\documentstyle[aps,epsf,preprint]{revtex}
\newcommand{\lkk}{\left[}  \newcommand{\rkk}{\right]}
  
\newcommand{\omegamz}{\Omega_{\rm m0}}

\newcommand{\omegavz}{\Omega_{\rm v0}}
\newcommand{\solar}{\odot}
\begin{document}

\begin{center}
{\large\bf Probing the star formation epoch with Wilkinson Microwave
 Anisotropy Probe (WMAP) data and high-redshift QSOs
}
\end{center}
\vspace{3mm}

\begin{center}
{
 Jun'Ichi Yokoyama
\vskip 1mm
\sl{Department of Earth and Space Science,
Graduate School of Science,\\
Osaka University, Toyonaka 560-0043, Japan}}
\end{center}

\begin{abstract}

Using the values of the cosmological parameters obtained from 
Wilkinson Microwave
 Anisotropy Probe (WMAP) and the cosmological chemical clock
observed by high-redshift quasars, it is shown that star formation should have
 started by the redshift $z_F > 9.5$, and that $z_F$ is sensitive only to
 $\omegamz h^2$ in flat $\Lambda$CDM cosmology.

\end{abstract}

\newpage
\tighten
The first year data release of Wilkinson Microwave
 Anisotropy Probe (WMAP) has opened a new era of high precision
 cosmology (Bennet et al. 2003).
As for the values of the cosmological parameters it has basically
 confirmed that of the concordance model with significantly smaller
 error bars than before.  Together with the observation of nearly
 scale-invariant adiabatic fluctuations, the inflation-based flat
 $\Lambda$CDM model can be regarded as the primary candidate of
the standard evolution model of our Universe.

Ever since Hubble's discovery of cosmic expansion (Hubble 1929), 
who concluded that
the cosmic expansion rate was as large as $H_0 \simeq 500$km/s/Mpc, 
we encountered the cosmic age problem from time to time in the history
of modern cosmology.  The introduction of vacuum-like energy density with
negative pressure such as  cosmological constant $\Lambda$ as inferred from 
the magnitude-redshift relation of high-redshift type Ia supernovae
(Perlmutter et al.\ 1999; Riess et al.\ 1998)
 has
reconciled a relatively large Hubble parameter
$h=H_0/(100$km/s/Mpc)$\simeq 0.7$ 
(Freedmann et al.\ 2001)
with a fairly large cosmic age $t_0
\gtrsim 13$Gyr.  Indeed the cosmic age obtained by WMAP observations
supplemented by other cosmological observations, $t_0 =
13.7 \pm 0.2$Gyr with $h=0.71^{+0.04}_{-0.03}$, 
$\omegamz h^2=0.135^{+0.008}_{-0.009}$,
$\omegavz=\Omega_{\rm tot0}-\omegamz$, and  
$\Omega_{\rm tot0}=1.02\pm0.02$
are in good agreement
with astrophysical estimate of globular cluster age and cosmological 
nuclear chronology (Spergel et al. 2003).  
Here $\omegavz$ denotes the current value of the
 vacuum-like energy density or dark energy in unit of the critical
 density.

Thus as far as the current Universe is concerned we have indeed the
concordance of the cosmic age and various cosmological parameters
 at hand.  However,  the situation may change
drasitically
if we turn our attention to the
high-redshift Universe.  Indeed
with the improved accuracy of current values of the cosmological
parameters, we can calculate the cosmic age at redshift $z$, $t(z)$,
more accurately.  Comparing $t(z)$ with other measures of cosmic age at
high-redshift we can obtain important cosmological and astrophysical
information and constraints.

As for the latter we consider observation of Fe/Mg abundance
ratio probed by high-redshift quasars.  This quantity can  serve
 as a cosmic chemical clock due to the difference of the origin
(Hamann \& Ferland 1993). 
That is, Fe is predominantly produced by type Ia supernovae whose
precursor has a lifetime of
 $\sim$ 1 Gyr or larger, while Mg and other alpha
elements are produced from type II supernovae with much shorter time
scale.  Thus the ratio Fe/Mg at the redshift $z$ 
indicates the time elapsed between the star
formation epoch $z_F$ and $z$.  

Previously this measure was mainly used
to constrain the values of cosmological parameters with particular
emphasis on whether cosmological constant is nonvanishing
(Yoshii, Tsujimoto, \& Kawara 1998).
But at the new era of precision cosmology with WMAP and other
sophisticated observational tools we can extract useful information on
$z_F$ and chemical evolution models of the Universe as argued below.

Although the observational constraint of WMAP on the spatial curvature or 
the total energy density of the Universe still has a two percent
uncertainty, $\Omega_{\rm tot0}=1.02\pm0.02$, we assume $\Omega_{\rm
tot0}$ is equal to unity with much higher accuracy, because we have good
reasons to believe in standard inflation as argued above, which predicts
$|\Omega_{\rm tot0}-1|<Q\simeq 5\times 10^{-6}$ with $Q$ being the
quadrupole anisotropy of cosmic microwave background radiation
(Kashlinsky, Tkachev \& Frieman 1994).
Then, in the spatially flat Robsertson-Walker spacetime with the scale factor
 $a(t)$,
 cosmic age at
redshift $z$ is given by
\begin{eqnarray}
t\left( z \right) &=& \int {\frac{{da}}{{\dot a}}}  
= \frac{1}{{H_0 }}\int_z^\infty  {\frac{{dz}}{{(1 + z)E(z)}}}, \nonumber \\
E(z) &=& \left[ {\Omega _{{\rm{m0}}} (1 + z)^3 
  + \Omega _{{\rm{v0}}} f(z)} \right]^{{1 \mathord{\left/
 {\vphantom {1 2}} \right.
 \kern-\nulldelimiterspace} 2}} , \label{age}
\end{eqnarray}
where $f(z)$ is a function which specifies the evolution of dark energy
component.  If it has a constant index of equation of state, $w_v \equiv
\rho_v/p_v$, where $\rho$ and $p$ denote energy density and pressure
respectively, one finds $f(z)=(1+z)^{3(1+w_v)}$.  The pure cosmological
constant corresponds to $w_v=-1$.  Below we mainly consider this case
because it is theoretically well-motivated (Yokoyama 2002a,b) and
observationally supported among many
 other candidates of dark energy (Yokoyama 2003).  In the
above expression we have neglected radiation component, which is a good
approximation in the range of redshifts we consider.

Several authors have calculated time evolution of Fe/Mg ratio using a
chemical evolution model.  For example, Yoshii, Tsujimoto \& Kawara
(1998) calculated its evolution based on the model of Yoshii, Tsujimoto
\& Nomoto (1996) with the following three basic ingredients.  
The first one is the star formation rate $C(t)$ parametrized as 
$C(t)=\nu_k\lkk f_g(t)\rkk^k$ where $f_g(t)$ is the gas fraction and
$\nu_k$ and $k$ are parameters.  As the standard model they choose $k=1$
and $\nu_{k=1}=7.6{\rm Gyr}^{-1}$ as inferred from metal abundance in
quasar host galaxies (Hamann \& Ferland 1993).  
The second is the initial stellar mass function
(IMF) which is assumed to be time-invariant with a spectrum
$\phi(m)dm \propto m^{-x}dm$ for $0.05M_{\solar}\leq m \leq 50M_{\solar}$
with $x=1.35$ based on Tsujimoto et al.\ (1997).  The third and the most
important ingredient is the fraction, $A$, and the lifetime, $t_{Ia}$,
of the progenitors that eventually produce SNe Ia, and their time spread.
They adopt $A=0.055$ for $m=3-8M_{\solar}$ and $A=0$ outside this mass
range as a standard value.  $t_{Ia}$ is obtained from the break in
[O/Fe] observed at [Fe/H]$\sim -1$ in the solar neighborhood 
(Yoshii, Tsujimoto \& Nomoto 1996)
and taken
as $t_{Ia}=1.5$ Gyr, while its spread function, $g(t_{Ia})$, is 
modeled as a power-law $g(t_{Ia})\propto t_{Ia}^\gamma$.  In practice,
they take $\gamma=0$ and $g(t_{Ia})\neq 0$ for $t_{Ia}=1-3$ Gyr, again to
match the observational features of [O/Fe] and [Fe/H].

As for the observation, Yoshii, Tsujimoto and Kawara used UV-optical
spectrum of QSO B1422+231 at $z=3.62$ and found FeII(UV+opt)/MgII
$\lambda\lambda$ 2796, 2804 flux ratio to be $12.2\pm 3.9$.  Assuming this
ratio traces the abundance ratio Fe/Mg, they obtained 
[Mg/Fe]$=-0.61^{+0.12+0.16}_{-0.30-0.12}$, which, according to
their standard evolution model, implies that the time elapsed from $z_F$ to
$z=3.62$ is given by
\begin{equation}
 \Delta t_{z=3.62}\equiv t(z=3.62)-t(z_F)=1.50~{\rm Gyr}.
\end{equation}
They have also considered possible modification of their standard model
and concluded that the lower bound is 1.3 Gyr, namely, 
$\Delta t_{z=3.62}\geq 1.3$ Gyr, which may be realized for a higher value of $A
\gtrsim 0.08$.  As is seen in Table 1 of Yoshii, Tsujimoto and Kawara
(1998), modification of other model parameters tends to increase 
$\Delta t$.

This quasar is by no means the only QSO whose flux ratio FeII/MgII has
been observed.  Thompson, Hall and Elston (1999) analyzed spectra of 12
high-redshift quasars to probe evolution of FeII/MgII ratio.  As a
result they found no decline in the iron abundance even at $z=4.47$ and
concluded that about 1Gyr had passed from $z_F$ to $z=4.47$.
Dietrich et al.\ (2002), on the other hand, observed six high-redshift
QSOs with $z=3310-3.488$ and basically confirmed the conclusion of
Yoshii, Tsujimoto and Kawara (1998), namely, 
$\Delta t_{z\approx 3.4}\approx 1.5$ Gyr.

Comparing these values with the formula (\ref{age}) calculated with the
improved values of cosmological parameters obtained with WMAP, we can
find constraint on the epoch of star formation $z_F$.
Specifically we consider the three possible tests on $z_F$ 
as argued above, to draw contours of $z_F$ corresponding to
$\Delta t_{z=3.62}= 1.3$ Gyr, $\Delta t_{z=3.62}= 1.5$ Gyr
and $\Delta t_{z=4.77}= 1.0$ Gyr.

Figures 1-3 depict these contour on an $\omegamz - H_0$ plane.  As is
seen there the condition $\Delta t_{z=3.62}= 1.5$ Gyr is more stringent
than that $\Delta t_{z=4.77}= 1.0$ Gyr.  For the most likely value of
the cosmological parameters obtained by WMAP, $h=0.71$ and
$\omegamz=0.268$ under the assumption that $\omegavz=1-\omegamz$, we
find $z_F= 10.13$ from $\Delta t_{z=3.62}= 1.3$ Gyr, 
$z_F= 14.98$ from $\Delta t_{z=3.62}= 1.5$ Gyr, and
$z_F= 11.91$ from $\Delta t_{z=4.77}= 1.0$ Gyr.  Thus significant star
formation before $z_F > 10$ is inferred.
In principle we could yield even more stringent constraint than
$z_F=14.98$, if we applied Yoshii et al's model (1998) to Thompson et al's
data (1999), but we remain conservative here, allowing the room for
uncertainty in chemical evolution models.
 
More interesting are contour maps depicted on a $\omegamz h^2-H_0$ plane,
which are shown in figs.\ 4-6.  As is seen there in the range of
cosmological parameters under consideration $z_F$ is insensitive to
$H_0$ and depends only on $\omegamz h^2$.  Thus for $\omegamz
h^2=0.135^{+0.008}_{-0.009}$, we obtain
$z_F=10.1_{-0.6}^{+0.7}$  from $\Delta t_{z=3.62}= 1.3$ Gyr, 
$z_F=15.0_{-1.7}^{+1.9}$ from $\Delta t_{z=3.62}= 1.5$ Gyr, and
$z_F=11.9 \pm0.7$ from  $\Delta t_{z=4.77}= 1.0$ Gyr.
We may therefore conclude $z_F > 9.5$ in the flat $\Lambda$CDM Universe
with the cosmological parameters obtained by combination of WMAP and
other measures.

Finally we briefly mention the case dark energy has a softer equation of
state $w>-1$ than the cosmological constant.  For a constant coefficient
$w$, WMAP has obtained a constraint $w<-0.78$ (Spergel et al. 2003).  
Dark energy with a
softer equation of state tends to require star formation at higher
redshift, although the change is modest.  For example, for $w=-0.8$,
$h=0.71$, and $\omegamz=0.268$ we find $z_F=10.36,$ 15.57, and 12.10,
respectively instead of $z_F=10.13,$ 14.98, and 11.91 for the case of 
the cosmological constant as discussed above. 

In conclusion we have pointed out that in the new era of precision
cosmology we can obtain some useful information on the evolution of the
high-redshift Universe from the cosmological chemical clock rather than
constraining current values of cosmological parameters with them.  
We have shown that the epoch of
star formation $z_F$ should satisfy $z_F>9.5$ and that it
is sensitive only to $\omegamz h^2$ in the context of the
standard flat $\Lambda$CDM cosmology. This constraint would bring about
interesting implication to formation of galaxies that host quasars.

\vskip 1cm 
This work was partially supported by the JSPS Grant-in-Aid for
Scientific Research No.\ 13640285.

\begin{figure}
\begin{center}
\leavevmode
\epsfxsize=7.05cm
\epsfbox{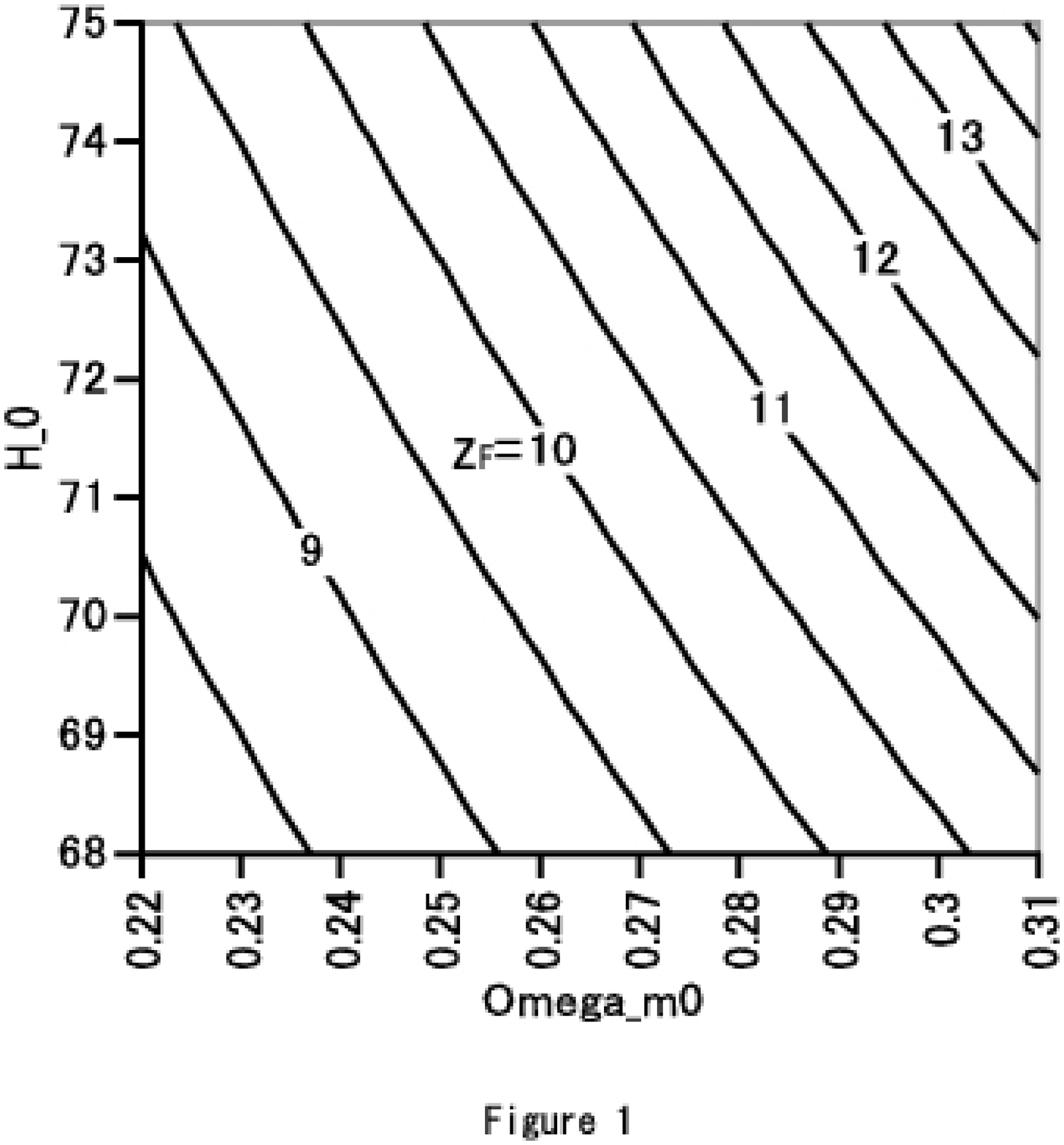}
\caption{Contour map of $z_F$ which satisfies $\Delta t_{z=3.62}= 1.3$
 Gyr as a function of $\omegamz$ and $H_0$. }
\end{center}
\end{figure}
\begin{figure}
\begin{center}
\leavevmode
\epsfxsize=6.9cm
\epsfbox{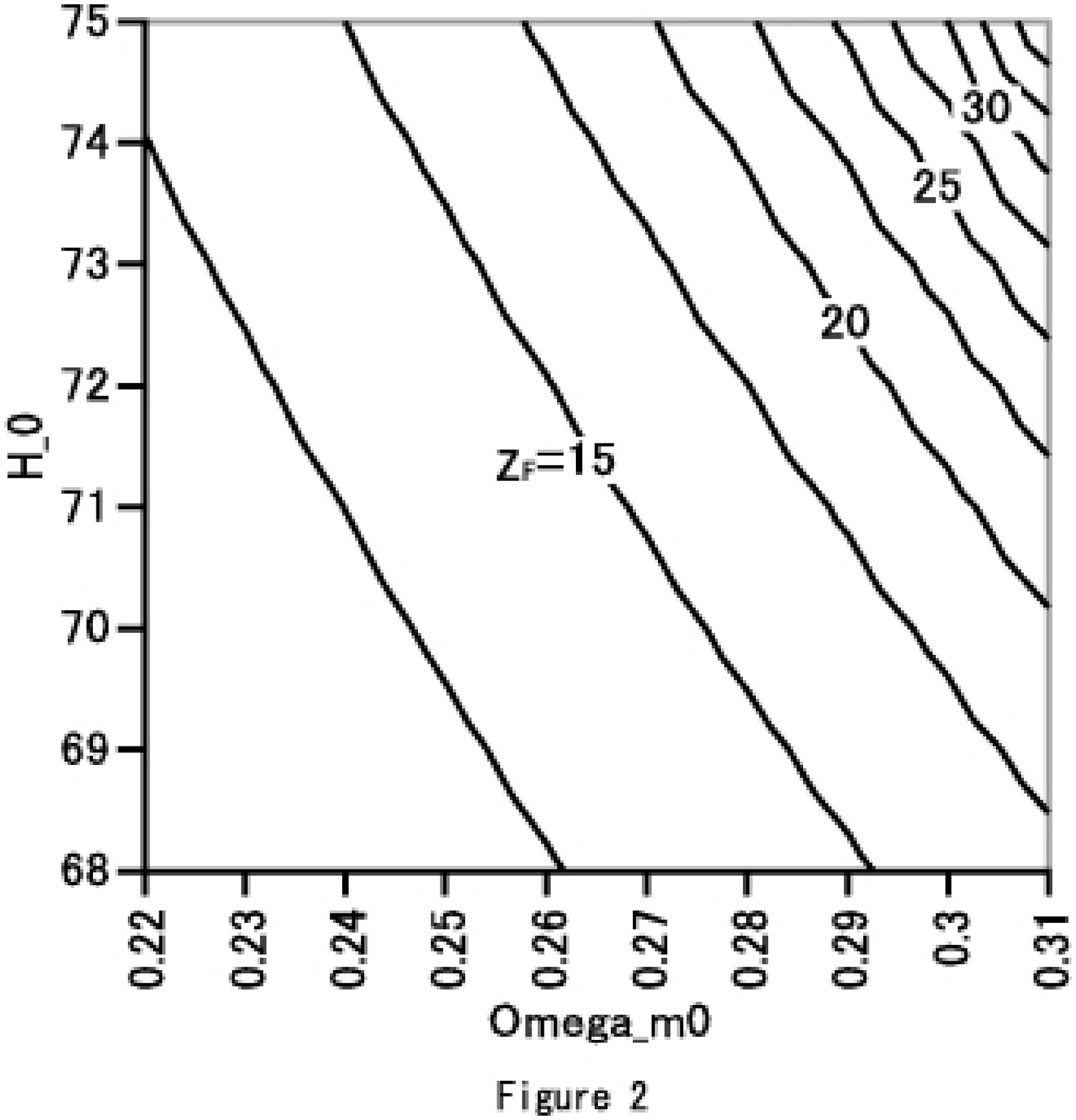}
\caption{Contour map of $z_F$ which satisfies $\Delta t_{z=3.62}= 1.5$
 Gyr as a function of $\omegamz$ and $H_0$. }
\end{center}
\end{figure}\begin{figure}
\begin{center}
\leavevmode
\epsfxsize=7.0cm
\epsfbox{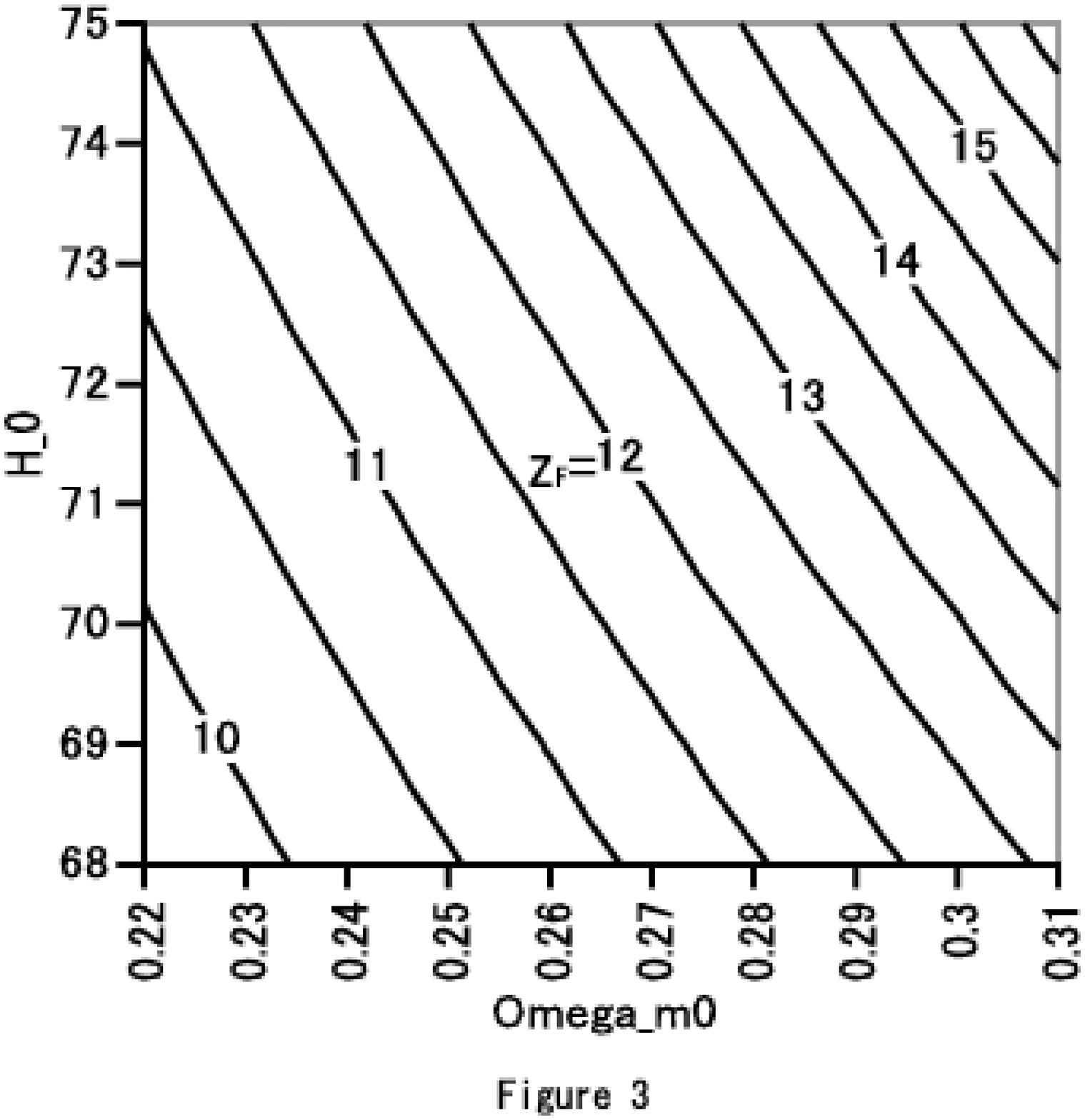}
\caption{Contour map of $z_F$ which satisfies $\Delta t_{z=4.77}= 1.0$
 Gyr as a function of $\omegamz$ and $H_0$. }
\end{center}
\end{figure}

\begin{figure}
\begin{center}
\leavevmode
\epsfxsize=7.05cm
\epsfbox{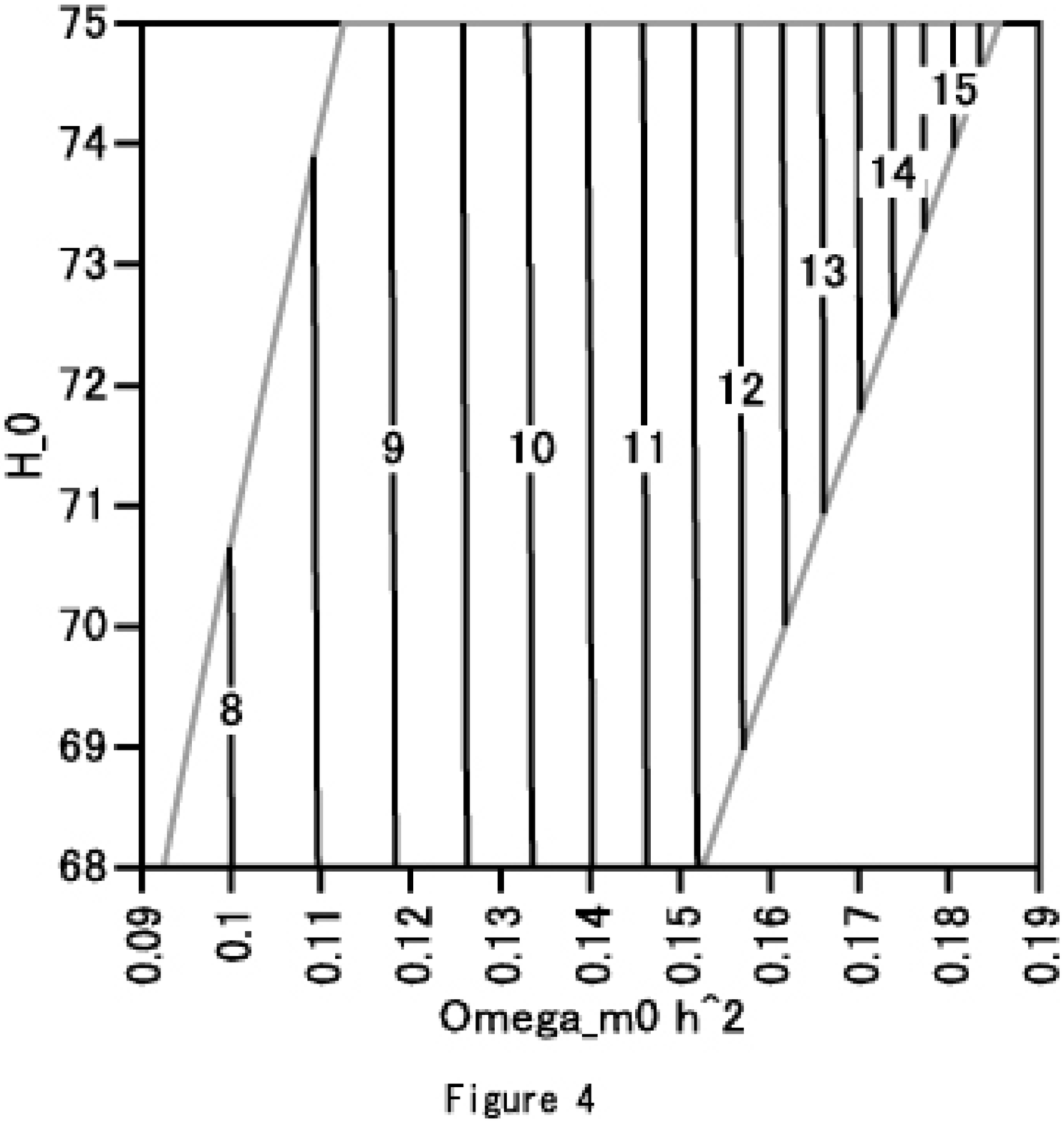}
\caption{Contour map of $z_F$ which satisfies $\Delta t_{z=3.62}= 1.3$
 Gyr as a function of $\omegamz h^2$ and $H_0$. }
\end{center}
\end{figure}
\begin{figure}
\begin{center}
\leavevmode
\epsfxsize=6.9cm
\epsfbox{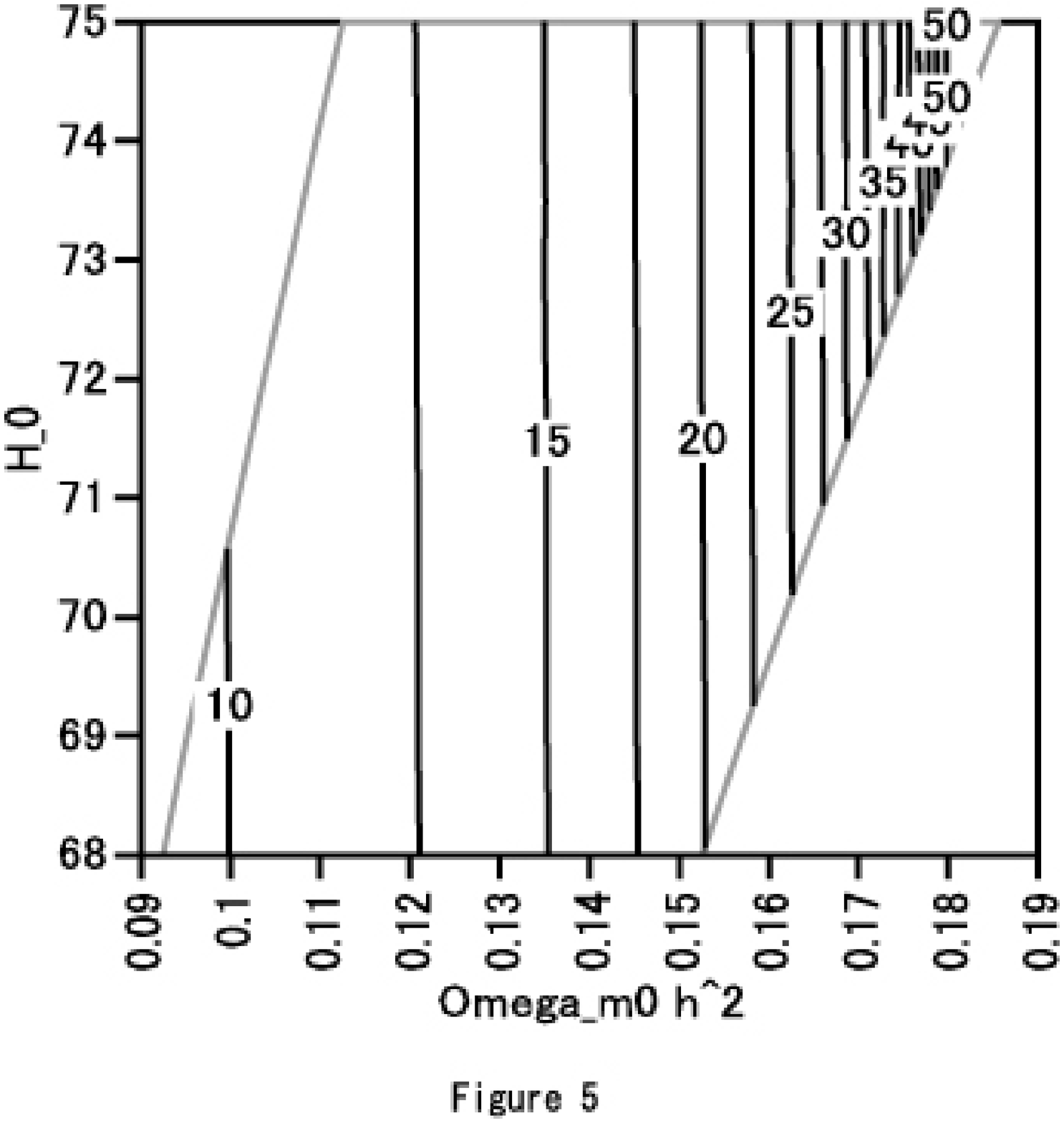}
\caption{Contour map of $z_F$ which satisfies $\Delta t_{z=3.62}= 1.5$
 Gyr as a function of $\omegamz h^2$ and $H_0$. }
\end{center}
\end{figure}\begin{figure}
\begin{center}
\leavevmode
\epsfxsize=7.0cm
\epsfbox{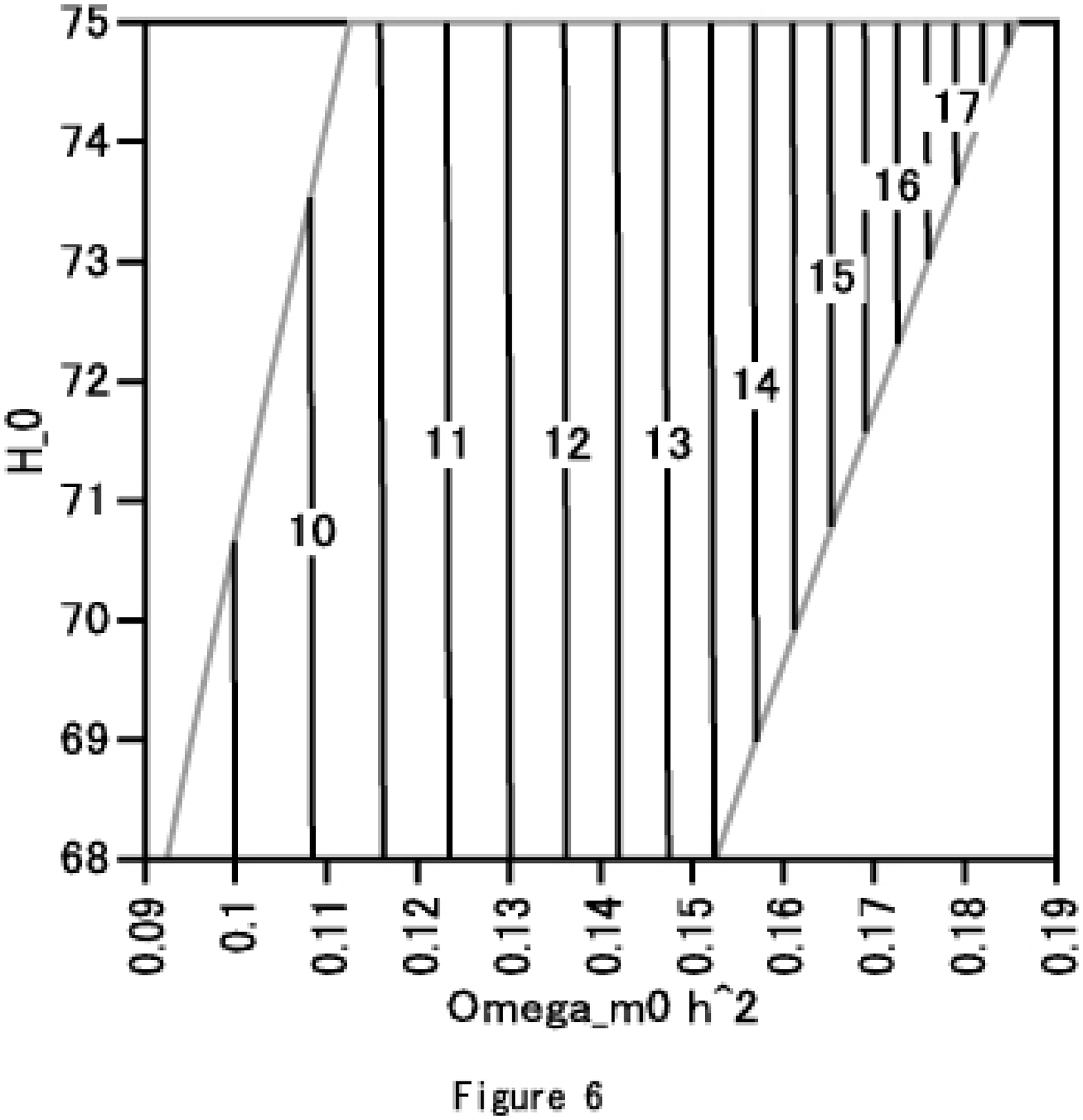}
\caption{Contour map of $z_F$ which satisfies $\Delta t_{z=4.77}= 1.0$
 Gyr as a function of $\omegamz h^2$ and $H_0$. }
\end{center}
\end{figure}

\begin{description}
\item{}Bennett, C.L., Halpern, M., Hinshaw, G., Jarosik, N., Kogut, A., 
Limon, M., Meyer, S.S., Page, L., Spergel, D.N. et al., 2003,
	   astro-ph/0302207
\item{}Dietrich, M., Appenzeller, I., Vestergaard, M., and Wagner, S.J.,
2002, ApJ, {\bf 564}, 581
\item{}Freedman, W.L., Madore, B.F., Gibson, B.K., Ferrarese, L.,
	   Kelson, D.D., Sakai, S., Mould, J.R., Kennicutt, R.C.\
 et al., 2001, ApJ, {\bf 553}, 47
\item{}Hamann, F.\ and Ferland, G., 1993, ApJ, {\bf 418}, 11
\item{}Hubble, E.P, 1929, Proc.\ Nat.\ Acad.\ Sci, {\bf 15}, 168
\item{}Kashlinsky, A., Tkachev, I.I., Frieman, J., 1994, Phys.\ Rev.\
	   Lett., {\bf 73}, 1582
\item{}Perlmutter, S., Aldering, G., Goldhaber, G., Knop, R.A., 
Nugent, P., Castro, P.G., Deustua, S., Fabbro, S., Goobar, A., 
\ et al., 1999, ApJ, {\bf 517}, 565
\item{}Riess, A.G., Nugent, P.E., Gilliland, R.L., Schmidt, B.P., Tonry,
	   J., Dickinson, M., Thompson, R.I., Budavari, T.S., 
\ et al., 1998 AJ, {\bf 116}, 1009
\item{}Spergel, D.N.,  Verde, L.,  Peiris, H.V.,  Komatsu, E., 
Nolta, M.R.,   Bennett, C.L.,   Halpern, M.,  Hinshaw, G., et al.,
2003, astro-ph/0302209
\item{}Thompson, K.L., Hill, G.J., Elston, R., 1999, ApJ, {\bf 515}, 487
\item{}Tsujimoto, T., Yoshii, Y., Nomoto, K., Matteucci, F., Thielemann,
	   F.-K., \& Hashimoto, M., 1997, ApJ, {\bf 483}, 228
\item{}Yoshii, Y., Tsujimoto, T., and Nomoto, K., 1996, ApJ, {\bf 462}, 266
\item{}Yoshii, Y., Tsujimoto, T., and Kawara, K., 1998, ApJL, {\bf 507}, L113
\item{}Yokoyama, J.\ 2002a, Phys.\ Rev.\ Lett., {\bf 88}, 151302 
\item{}Yokoyama, J.\ 2002b, Int.\ J.\ Mod.\ Phys., {\bf D11}, 1603 
\item{}Yokoyama, J.\ 2003, In Proc.\ 12th workshop on general relativity
	   and gravitation, eds.\ Y.\ Eriguchi and M.\
	   Shibata. (University of Tokyo)
\end{description}

\end{document}